\documentclass[10pt,a4paper,onecolumn]{article}
\usepackage{marginnote}
\usepackage{graphicx}
\usepackage{xcolor}
\usepackage{authblk,etoolbox}
\usepackage{titlesec}
\usepackage{calc}
\usepackage{tikz}
\usepackage{hyperref}
\hypersetup{colorlinks,breaklinks=true,
            urlcolor=[rgb]{0.0, 0.5, 1.0},
            linkcolor=[rgb]{0.0, 0.5, 1.0}}
\usepackage{caption}
\usepackage{tcolorbox}
\usepackage{amssymb,amsmath}
\usepackage{ifxetex,ifluatex}
\usepackage{seqsplit}
\usepackage{xstring}

\usepackage{float}
\let\origfigure\figure
\let\endorigfigure\endfigure
\renewenvironment{figure}[1][2] {
    \expandafter\origfigure\expandafter[H]
} {
    \endorigfigure
}

\usepackage{fixltx2e} % provides \textsubscript
\usepackage[
  backend=biber,
]{biblatex}
\bibliography{paper.bib}

\let\textttOrig=\texttt
\def\texttt#1{\expandafter\textttOrig{\seqsplit{#1}}}
\renewcommand{\seqinsert}{\ifmmode
  \allowbreak
  \else\penalty6000\hspace{0pt plus 0.02em}\fi}

\makeatletter
\let\href@Orig=\href
\def\href@Urllike#1#2{\href@Orig{#1}{\begingroup
    \def\Url@String{#2}\Url@FormatString
    \endgroup}}
\def\href@Notdoi#1#2{\def\tempa{#1}\def\tempb{#2}%
  \ifx\tempa\tempb\relax\href@Urllike{#1}{#2}\else
  \href@Orig{#1}{#2}\fi}
\def\href#1#2{%
  \IfBeginWith{#1}{https://doi.org}%
  {\href@Urllike{#1}{#2}}{\href@Notdoi{#1}{#2}}}
\makeatother

\newlength{\cslhangindent}
\newlength{\csllabelwidth}
\newenvironment{CSLReferences}[2] % #1 hanging-ident, #2 entry spacing
 {% don't indent paragraphs
  \setlength{\parindent}{0pt}
  \ifodd #1 \everypar{\setlength{\hangindent}{\cslhangindent}}\ignorespaces\fi
  \ifnum #2 > 0
  \setlength{\parskip}{#2\baselineskip}
  \fi
 }%
 {}
\usepackage{calc}

\usepackage[top=3.5cm, bottom=3cm, right=1.5cm, left=1.0cm,
            headheight=2.2cm, reversemp, includemp, marginparwidth=4.5cm]{geometry}

\titleformat{\section}
  {\normalfont\sffamily\Large\bfseries}
  {}{0pt}{}
\titleformat{\subsection}
  {\normalfont\sffamily\large\bfseries}
  {}{0pt}{}
\titleformat{\subsubsection}
  {\normalfont\sffamily\bfseries}
  {}{0pt}{}
\titleformat*{\paragraph}
  {\sffamily\normalsize}

\usepackage{fancyhdr}
\makeatletter
\let\ps@plain\ps@fancy
\fancyheadoffset[L]{4.5cm}
\fancyfootoffset[L]{4.5cm}

\definecolor{linky}{rgb}{0.0, 0.5, 1.0}

\newtcolorbox{repobox}
   {colback=red, colframe=red!75!black,
     boxrule=0.5pt, arc=2pt, left=6pt, right=6pt, top=3pt, bottom=3pt}

\newcommand{\ExternalLink}{%
   \tikz[x=1.2ex, y=1.2ex, baseline=-0.05ex]{%
       \begin{scope}[x=1ex, y=1ex]
           \clip (-0.1,-0.1)
               --++ (-0, 1.2)
               --++ (0.6, 0)
               --++ (0, -0.6)
               --++ (0.6, 0)
               --++ (0, -1);
           \path[draw,
               line width = 0.5,
               rounded corners=0.5]
               (0,0) rectangle (1,1);
       \end{scope}
       \path[draw, line width = 0.5] (0.5, 0.5)
           -- (1, 1);
       \path[draw, line width = 0.5] (0.6, 1)
           -- (1, 1) -- (1, 0.6);
       }
   }

\patchcmd{\@maketitle}{center}{flushleft}{}{}
\patchcmd{\@maketitle}{center}{flushleft}{}{}
\patchcmd{\@maketitle}{\LARGE}{\LARGE\sffamily}{}{}
\def\maketitle{{%
  
  \AB@maketitle}}
\makeatletter
\renewcommand\AB@affilsepx{ \protect\Affilfont}
\renewcommand\AB@affilnote[1]{{\bfseries #1}\hspace{3pt}}
\renewcommand{\affil}[2][]%
   {\newaffiltrue\let\AB@blk@and\AB@pand
      \if\relax#1\relax\def\AB@note{\AB@thenote}\else\def\AB@note{#1}%
        \setcounter{Maxaffil}{0}\fi
        \begingroup
        \let\href=\href@Orig
        \let\texttt=\textttOrig
        \let\protect\@unexpandable@protect
        \def\thanks{\protect\thanks}\def\footnote{\protect\footnote}%
        \@temptokena=\expandafter{\AB@authors}%
        {\def\\{\protect\\\protect\Affilfont}\xdef\AB@temp{#2}}%
         \xdef\AB@authors{\the\@temptokena\AB@las\AB@au@str
         \protect\\[\affilsep]\protect\Affilfont\AB@temp}%
         \gdef\AB@las{}\gdef\AB@au@str{}%
        {\def\\{, \ignorespaces}\xdef\AB@temp{#2}}%
        \@temptokena=\expandafter{\AB@affillist}%
        \xdef\AB@affillist{\the\@temptokena \AB@affilsep
          \AB@affilnote{\AB@note}\protect\Affilfont\AB@temp}%
      \endgroup
       \let\AB@affilsep\AB@affilsepx
}
\makeatother

\renewcommand\Affilfont{\sffamily\small\mdseries}
\ifnum 0\ifxetex 1\fi\ifluatex 1\fi=0 % if pdftex
  \usepackage[T1]{fontenc}
  \usepackage[utf8]{inputenc}

\else % if luatex or xelatex
  \ifxetex
    \usepackage{mathspec}
    \usepackage{fontspec}

  \else
    \usepackage{fontspec}
  \fi
  \defaultfontfeatures{Ligatures=TeX,Scale=MatchLowercase}

\fi
\IfFileExists{upquote.sty}{\usepackage{upquote}}{}
\IfFileExists{microtype.sty}{%
\usepackage{microtype}
\UseMicrotypeSet[protrusion]{basicmath} % disable protrusion for tt fonts
}{}

\usepackage{hyperref}
\hypersetup{unicode=true,
            pdftitle={DBSP\_DRP: A Python package for automated spectroscopic data reduction of DBSP data},
            pdfborder={0 0 0},
            breaklinks=true}
\let\addcontentslineOrig=\addcontentsline
\def\addcontentsline#1#2#3{\bgroup
  \let\texttt=\textttOrig\addcontentslineOrig{#1}{#2}{#3}\egroup}
\let\markbothOrig\markboth
\def\markboth#1#2{\bgroup
  \let\texttt=\textttOrig\markbothOrig{#1}{#2}\egroup}
\let\markrightOrig\markright
\def\markright#1{\bgroup
  \let\texttt=\textttOrig\markrightOrig{#1}\egroup}

\usepackage{graphicx,grffile}
\makeatletter
\def\maxwidth{\ifdim\Gin@nat@width>\linewidth\linewidth\else\Gin@nat@width\fi}
\def\maxheight{\ifdim\Gin@nat@height>\textheight\textheight\else\Gin@nat@height\fi}
\makeatother
\setkeys{Gin}{width=\maxwidth,height=\maxheight,keepaspectratio}
\IfFileExists{parskip.sty}{%
\usepackage{parskip}
}{% else
\setlength{\parindent}{0pt}
\setlength{\parskip}{6pt plus 2pt minus 1pt}
}
\ifx\paragraph\undefined\else
\let\oldparagraph\paragraph
\renewcommand{\paragraph}[1]{\oldparagraph{#1}\mbox{}}
\fi
\ifx\subparagraph\undefined\else
\let\oldsubparagraph\subparagraph
\renewcommand{\subparagraph}[1]{\oldsubparagraph{#1}\mbox{}}
\fi

\title{DBSP\_DRP: A Python package for automated spectroscopic data
reduction of DBSP data}

        \author[1,2]{Milan Sharma Mandigo-Stoba}
          \author[2]{Christoffer Fremling}
          \author[2]{Mansi M. Kasliwal}
    
      \affil[1]{Schmidt Academy of Software Engineering, California
Institute of Technology}
      \affil[2]{Division of Physics, Mathematics and Astronomy,
California Institute of Technology}
  \date{\vspace{-7ex}}

\begin{document}
\maketitle

\marginpar{

  \begin{flushleft}
  %\hrule
  \sffamily\small

  {\bfseries DOI:} \href{PENDING}{\color{linky}{Pending DOI}}

  \vspace{2mm}

  {\bfseries Software}
  \begin{itemize}
    \setlength\itemsep{0em}
    \item \href{https://github.com/openjournals/joss-reviews/issues/3511}{\color{linky}{Review}} \ExternalLink
    \item \href{https://github.com/finagle29/DBSP_DRP/}{\color{linky}{Repository}} \ExternalLink
    \item \href{PENDING}{\color{linky}{Pending Archive}} \ExternalLink
  \end{itemize}

  \vspace{2mm}

  \par\noindent\hrulefill\par

  \vspace{2mm}

  {\bfseries Editor:} \href{http://example.com}{Pending
Editor} \ExternalLink \\
  \vspace{1mm}
    \vspace{2mm}

  {\bfseries Submitted:} 19 July 2021

  \vspace{2mm}
  {\bfseries License}\\
  Authors of papers retain copyright and release the work under a Creative Commons Attribution 4.0 International License (\href{http://creativecommons.org/licenses/by/4.0/}{\color{linky}{CC BY 4.0}}).

  \end{flushleft}
}

\hypertarget{summary}{%
\section{Summary}\label{summary}}

In astronomy, the spectrum of light emitted from astrophysical sources
is of great use, allowing astronomers to classify objects and measure
their properties. To measure the spectrum, astronomers use
spectrographs, which use dispersive elements to split the incoming light
into its constituent wavelengths, and then image this dispersed light
with a detector, most commonly a CCD. But to do science with the
spectrum, the 2D image in pixel coordinates taken by the CCD must be
converted into a 1D spectrum of flux vs.~wavelength. To increase the
signal-to-noise ratio, astronomers can take multiple exposures of the
same object and coadd their 1D spectra to reveal faint absorption lines
or increase the precision with which an important emission line can be
measured. Many spectrographs have multiple paths that light can go
through, and multiple detectors, each measuring a particular part of the
spectrum, to increase the wavelength range that can be captured in a
single exposure, or to allow the high resolution observation of distinct
wavelength ranges. If two detectors cover an overlapping region, caused
by partial reflectance of a dichroic (wavelength-dependent beam
splitter), then the spectra from each detector need to be spliced
together, combining the light collected by each detector. This process
of converting 2D CCD images into 1D spectra is called data reduction.

DBSP\_DRP is a python package that provides fully automated data
reduction of data taken by the Double Spectrograph (DBSP) at the
200-inch Hale Telescope at Palomar Observatory
(\protect\hyperlink{ref-Okeux5cux26Gunn1982}{Oke \& Gunn, 1982}). The
underlying data reduction functionality to extract 1D spectra, perform
flux calibration and correction for atmospheric absorption, and coadd
spectra together is provided by PypeIt
(\protect\hyperlink{ref-Prochaska2020}{Prochaska et al., 2020}). The new
functionality that DBSP\_DRP brings is in orchestrating the complex data
reduction process by making smart decisions so that no user input is
required after verifying the correctness of the metadata in the raw FITS
files in a table-like GUI. Though the primary function of DBSP\_DRP is
to autmatically reduce an entire night of data without user input, it
has the flexibility for astronomers to fine-tune the data reduction with
GUIs for manually identifying the faintest objects, as well as exposing
the full set of PypeIt parameters to be tweaked for users with
particular science needs. DBSP\_DRP also handles some of the occasional
quirks specific to DBSP, such as swapping FITS header cards, adding (an)
extra null byte/s to FITS files making them not conform to the FITS
specification, and not writing the coordinates of the observation to
file. Additionally, DBSP\_DRP contains a quicklook script for making
real-time decisions during an observing run, and can open a GUI
displaying a minimally reduced exposure in under 15 seconds. Docker
containers are available for ease of deploying DBSP\_DRP in its
quicklook configuration (without some large atmospheric model files) or
in its full configuration.

\hypertarget{statement-of-need}{%
\section{Statement of Need}\label{statement-of-need}}

Palomar Observatory, located near San Diego, CA, is a multinational
observatory with a broad user base. Users come from large and small
institutions, and their observing experience ranges from novice to
expert. One responsibility for serving such a diverse user base is to
provide software data reduction pipelines for the most frequently used
instruments, such as the Palomar Double Spectrograph (DBSP). Although
DBSP was commissioned in 1982, it remains the workhorse instrument of
the 200'' Hale Telescope. It is used on 42\% of the nights in a year,
comprising nearly all of the valuable ``dark'' (moonless) time. In
previous years, standard astronomical practice left the data reduction
up to the user. However, attitudes in instrument building have shifted
since DBSP was built. The pipeline is now considered an indispensable
component of the astronomical instrument. In fact, the difference
between a good pipeline and a great pipeline means the difference
between counting some of the photons vs.~counting all of the photons.

Spectroscopy is a severe bottleneck in time-domain astronomy; currently
less than 10\% of discoveries are spectroscopically classified. Without
a pipeline, data reduction is a difficult process and the standard
method without a pipeline is to use IRAF, a 35 year old program on which
development and maintenance was discontinued in 2013 and whose use is
discouraged by many in the field e.g. Ogaz \& Tollerud
(\protect\hyperlink{ref-Ogaz2018}{2018}). Needless to say, data
reduction sans pipeline is extremely time-consuming. There is a clear
need for a modern and stable automated data reduction pipeline for DBSP.

During observing runs, one would like to be able to quickly inspect data
as it is taken, in order to ensure that it is of sufficient quality to
do the desired science with. For objects whose brightness may have
changed between a previous observation and the observing run, the
observer may have uncertainties regarding how long of an exposure is
needed to produce quality data. For very faint objects or objects in
crowded fields, the observer may not even be sure that the telescope is
pointed at the right object! A quicklook functionality, that can do a
rudimentary reduction to correct for instrumental signatures and
subtract light from the sky, revealing the spectra of the objects
observed, can answer questions of exposure time and whether the object
observed is the right one.

DBSP\_DRP is currently being used by the ZTF Bright Transient Survey
(\protect\hyperlink{ref-Fremling2020}{Fremling et al., 2020};
\protect\hyperlink{ref-Perley2020}{Perley et al., 2020}), the ZTF Census
of the Local Universe (\protect\hyperlink{ref-De2020}{De et al., 2020}),
and a program investigating ZTF Superluminous Supernovae (\protect\hyperlink{ref-Lunnan2020}{Lunnan et al., 2020}; Chen et al., in preparation). Ravi et al.
(\protect\hyperlink{ref-Ravi2021arXiv}{2021}) is the first (known)
publication that used DBSP\_DRP for data reduction. The development of
DBSP\_DRP also lays the groundwork towards a fully automated pipeline
for the Next Generation Palomar Spectrograph that is planned to be
deployed on the Palomar 200-inch Hale Telescope in 2022.

\hypertarget{acknowledgements}{%
\section{Acknowledgements}\label{acknowledgements}}

M.S.M.-S. acknowledges funding from the Schmidt Academy of Software
Engineering, which is supported by the generosity of Eric and Wendy
Schmidt by recommendation of the Schmidt Futures program.

We thank the following members of the time domain astronomy group at
Caltech for beta-testing and providing valuable feedback during the
development of this pipeline: Andy Tzanidakis, Lin Yan, Aishwarya
Dahiwale, Yuhan Yao, Yashvi Sharma, and Igor Andreoni.

M.S.M.-S. is extremely grateful to the welcoming, friendly, and helpful team
of developers on the PypeIt team, without whom this package would not
exist.

\hypertarget{references}{%
\section*{References}\label{references}}
\addcontentsline{toc}{section}{References}

\hypertarget{refs}{}
\begin{CSLReferences}{1}{0}
\leavevmode\vadjust pre{\hypertarget{ref-De2020}{}}%
De, K., Kasliwal, M. M., Tzanidakis, A., Fremling, U. C., Adams, S.,
Aloisi, R., Andreoni, I., Bagdasaryan, A., Bellm, E. C., Bildsten, L.,
Cannella, C., Cook, D. O., Delacroix, A., Drake, A., Duev, D., Dugas,
A., Frederick, S., Gal-Yam, A., Goldstein, D., \ldots{} Yao, Y. (2020).
{The Zwicky Transient Facility Census of the Local Universe. I.
Systematic Search for Calcium-rich Gap Transients Reveals Three Related
Spectroscopic Subclasses}. \emph{905}(1), 58.
\url{https://doi.org/10.3847/1538-4357/abb45c}

\leavevmode\vadjust pre{\hypertarget{ref-Fremling2020}{}}%
Fremling, C., Miller, A. A., Sharma, Y., Dugas, A., Perley, D. A.,
Taggart, K., Sollerman, J., Goobar, A., Graham, M. L., Neill, J. D.,
Nordin, J., Rigault, M., Walters, R., Andreoni, I., Bagdasaryan, A.,
Belicki, J., Cannella, C., Bellm, E. C., Cenko, S. B., \ldots{}
Kulkarni, S. R. (2020). {The Zwicky Transient Facility Bright Transient
Survey. I. Spectroscopic Classification and the Redshift Completeness of
Local Galaxy Catalogs}. \emph{The Astrophysical Journal}, \emph{895}(1),
32. \url{https://doi.org/10.3847/1538-4357/ab8943}

\leavevmode\vadjust pre{\hypertarget{ref-Lunnan2020}{}}%
Lunnan, R., Yan, L., Perley, D. A., Schulze, S., Taggart, K., Gal-Yam,
A., Fremling, C., Soumagnac, M. T., Ofek, E., Adams, S. M., Barbarino,
C., Bellm, E. C., De, K., Fransson, C., Frederick, S., Golkhou, V. Z.,
Graham, M. J., Hallakoun, N., Ho, A. Y. Q., \ldots{} Yao, Y. (2020).
{Four (Super)luminous Supernovae from the First Months of the ZTF
Survey}. \emph{The Astrophysical Journal}, \emph{901}(1), 61.
\url{https://doi.org/10.3847/1538-4357/abaeec}

\leavevmode\vadjust pre{\hypertarget{ref-Ogaz2018}{}}%
Ogaz, S., \& Tollerud, E. (2018). {Removing the Institute's Dependence
on IRAF (You can do it too!)}. \emph{STScI Newsletter}, \emph{35}(03).

\leavevmode\vadjust pre{\hypertarget{ref-Okeux5cux26Gunn1982}{}}%
Oke, J. B., \& Gunn, J. E. (1982). {An Efficient Low Resolution and
Moderate Resolution Spectrograph for the Hale Telescope}.
\emph{Publications of the Astronomical Society of the Pacific},
\emph{94}, 586. \url{https://doi.org/10.1086/131027}

\leavevmode\vadjust pre{\hypertarget{ref-Perley2020}{}}%
Perley, D. A., Fremling, C., Sollerman, J., Miller, A. A., Dahiwale, A.
S., Sharma, Y., Bellm, E. C., Biswas, R., Brink, T. G., Bruch, R. J.,
De, K., Dekany, R., Drake, A. J., Duev, D. A., Filippenko, A. V.,
Gal-Yam, A., Goobar, A., Graham, M. J., Graham, M. L., \ldots{} Yan, L.
(2020). {The Zwicky Transient Facility Bright Transient Survey. II. A
Public Statistical Sample for Exploring Supernova Demographics}.
\emph{The Astrophysical Journal}, \emph{904}(1), 35.
\url{https://doi.org/10.3847/1538-4357/abbd98}

\leavevmode\vadjust pre{\hypertarget{ref-Prochaska2020}{}}%
Prochaska, J. X., Hennawi, J. F., Westfall, K. B., Cooke, R. J., Wang,
F., Hsyu, T., Davies, F. B., Farina, E. P., \& Pelliccia, D. (2020).
PypeIt: The python spectroscopic data reduction pipeline. \emph{Journal
of Open Source Software}, \emph{5}(56), 2308.
\url{https://doi.org/10.21105/joss.02308}

\leavevmode\vadjust pre{\hypertarget{ref-Ravi2021arXiv}{}}%
Ravi, V., Law, C. J., Li, D., Aggarwal, K., Burke-Spolaor, S., Connor,
L., Lazio, T. J. W., Simard, D., Somalwar, J., \& Tendulkar, S. P.
(2021). {The host galaxy and persistent radio counterpart of FRB
20201124A}. \emph{arXiv e-Prints}, arXiv:2106.09710.
\url{https://arxiv.org/abs/2106.09710}

\end{CSLReferences}

\end{document}